  \documentclass[multi]{cambridge6Atight}
  \usepackage{natbib}

  \usepackage{rotating}
  \usepackage{floatpag}
  \rotfloatpagestyle{empty}

 \usepackage{mathptmx}

  \usepackage{makeidx}
  \makeindex

 \copyrightline{This material has been published in \textit{The Impact of Binaries on Stellar Evolution}, Beccari G. \& Boffin H.M.J. (Eds.). This version is free to view and download for personal use only. Not for re-distribution, re-sale or use in derivative works. \copyright\ 2018 Cambridge University Press.}

\newcommand\aap{A\&A}                
\newcommand\aapr{A\&ARv}             
\newcommand\aj{AJ}                   
\newcommand\apj{ApJ}                 
\newcommand\apjl{ApJ}                
\newcommand\araa{ARA\&A}             
\newcommand\mnras{MNRAS}             
\newcommand\nat{Nature}              
\newcommand\pasa{Publ. Astron. Soc. Australia}  
\newcommand\rmxaa{Rev. Mex. Astron. Astrofis.} 
    
\begin{document}


  \alphafootnotes
   \author[H\,Van Winckel]
    {Hans Van Winckel}
  \chapter{Binary post-AGB stars as tracers of stellar evolution}


  \contributor{Hans van Winckel
    \affiliation{Instituut voor Sterrenkunde, KU Leuven, Celestijnenaal 200D, 3001 Leuven, Belgium}}

 \begin{abstract}
 In this chapter the focus is on the properties of post-Asymptotic Giant Branch (post-AGB) stars in binary systems.
 Their Spectral Energy Distributions (SEDs) are very characteristic: they show a near-infrared excess, indicative of the presence of warm dust, while the central stars are too hot to be in a dust-production evolutionary phase. This allows for an efficient detection of binary post-AGB candidates.  It is now well established that the near-infrared excess is
produced by the inner rim of a stable dusty disc that surrounds the binary system. These
discs are scaled-up versions of protoplanetary discs and form a second generation of stable Keplerian discs. They are likely formed during a binary interaction process when the primary was on ascending the AGB.
I will summarise what we have learned from the observational properties
of these post-AGB binaries. The impact of the creation, lifetime and evolution of the circumbinary discs 
on the evolution of the system are yet to be fully understood. 
 
 \end{abstract}

\section{Introduction}
\label{sect:1}


The final evolution of low- and intermediate-mass stars (0.8 - 8 M$_{\odot}$) is a rapid transition from the Asymptotic
Giant Branch (AGB) over the post-AGB phase towards the Planetary Nebulae (PNe) stage. When most of
the stellar envelope is removed by an intense mass-loss rate (up to $10^{-4}$ M$_{\odot}$ /yr) at the end of the AGB, the star
enters the post-AGB phase of stellar evolution. During this phase the remaining stellar
envelope begins to shrink and the effective temperature starts to increase with almost constant luminosity. If the temperature increases on a
timescale shorter than the dispersion time of the matter previously ejected by the star, a PNe will
appear, as the result of the ionization of the circumstellar shell \citep{balick02, herwig05}. During the AGB to PNe transition, some objects show a resolved reflection nebula and these are referred to as proto-planetary nebula or pre-planetary nebula  \citep{sahai07}. These proto-planetary nebulae are only a subset of post-AGB stars, as the latter also includes objects that will not ionise their ejected matter as it is too dispersed. It is fair to say that the samples of proto-planetary nebulae and post-AGB stars in general, do not connect very well on evolutionary tracks with the AGB stars, nor with the PNe.  The diversity seen in individual examples is not well understood in the framework stellar evolution theory \cite[e.g.][]{vanwinckel03}.

One of the most important research questions regarding the final evolution of
low-and intermediate-mass stars, is the impact of binarity \citep[see recent review by][and references therein]{demarco17}. Also this book provides ample illustration of research results focussing on this question. 
Binary interaction alters the intrinsic properties of the evolved star (such as: pulsations, mass-loss,
dust-formation, circumstellar envelope morphology etc.) and plays a dominant role in determining
its ultimate fate. A plethora of peculiar objects ranging from the spectacular thermonuclear novae, supernovae
type Ia, sub-luminous supernovae, gravitational wave sources, etc. to less energetic systems such as sub-dwarf
B stars, barium stars, cataclysmic variables, bipolar PNe, etc., result from mass transfer in binary
stars. 

During evolution of the low- and intermediate-mass star, binary interactions will increase when one of the binary components evolves to giant dimensions. The dynamical behaviour of this interaction is determined by the balance between the Roche Lobe and the stellar radius of the red giant. Binaries with a main sequence separation of less than about two astronomical units, will be
tidally captured somewhere on the Red Giant Branch (RGB) or AGB. When unstable mass transfer ensues, a common envelope
(CE) event occurs, resulting in a dramatic shortening of the orbital separation. This might lead to a complete
stellar merger but, if this is avoided, the final period of these binaries after the giant phase is typically in the
order of days, depending on the envelope's binding energy. When the initial orbit is wider, the interaction is
thought to be via wind accretion or wind shaping and the primary component will evolve off the giant branch
with a very wide orbit. When the originally lower mass component evolves also onto the giant branch, interaction
may take place again, but this time with a compact companion. This theoretical scenario is borne out by
population-synthesis models \citep[e.g.][]{han02,han03}. The population-synthesis model normalised to the ellipsoidal binaries on the red giant phase, predicts the final period distribution of evolved binaries to
be bi-modal in which the CE channel results in short-period binaries (P $\sim$ 1 day) and the wind accretion channel
results in wider systems (P $>$ 1000 days) \citep{nie12}. The orbital periods of around 1000 days are least predicted and lie in the middle of this bi-modal distribution (see also Chapter by Pols in this book).

This is, however, in stark contrast to what is observed in post-AGB binaries and here we focus on the observed 
properties of these optically bright objects.
The reasons for this strong discrepancy between the observed and
predicted period distributions are twofold.  Firstly, a wide range of
binary interactions are not understood from first principles or are
even missing in the binary evolution models. The theoretical models
are therefore subject to many uncertainties like the efficiency of
envelope ejection \citep[e.g.][]{toonen13}; the
unknown physical description of the common-envelope phase
\citep[e.g.][]{izzard12, ivanova13}; the postulated increase of
the mass-loss prior to contact as to lower the envelope binding energy \citep[e.g.][]{chen11, abate13};  the badly understood impact of
radiation pressure on the shape of the classical Roche potential
\citep{dermine09}; the assumed mass transfer efficiency and its
orbital phase dependency. Secondly, there is also an observational
challenge as many parameters involved in predicting the outcome of the
binary evolution channels are not well constrained due to the lack of
observational data, where detailed studies of individual
objects prevail over systematic and time-consuming more complete
surveys.

In this chapter we focus on binary post-AGB stars and their observational properties. In Section~\ref{sect:SED} we focus on their Spectral Energy Distribution (SED) as this turned out to be very characteristic. We report on the orbital properties in Section~\ref{sect:orbits} and review all observational evidence for the presence of stable circumbinary discs as a common feature of all the binaries in Section~\ref{sect:discs}. Jets and their origin form the core of Section~\ref{sect:jets}. The impact of the feedback from the disc on the primary is given in Section~\ref{sect:depletion}. We end in Section~\ref{sect:conclusion} by naming a few challenges ahead to come to a better understanding of binary stellar evolution of low-and intermediate mass stars.

\section{SED} \label{sect:SED}

\begin{figure}
    \includegraphics[width=0.7\textwidth]{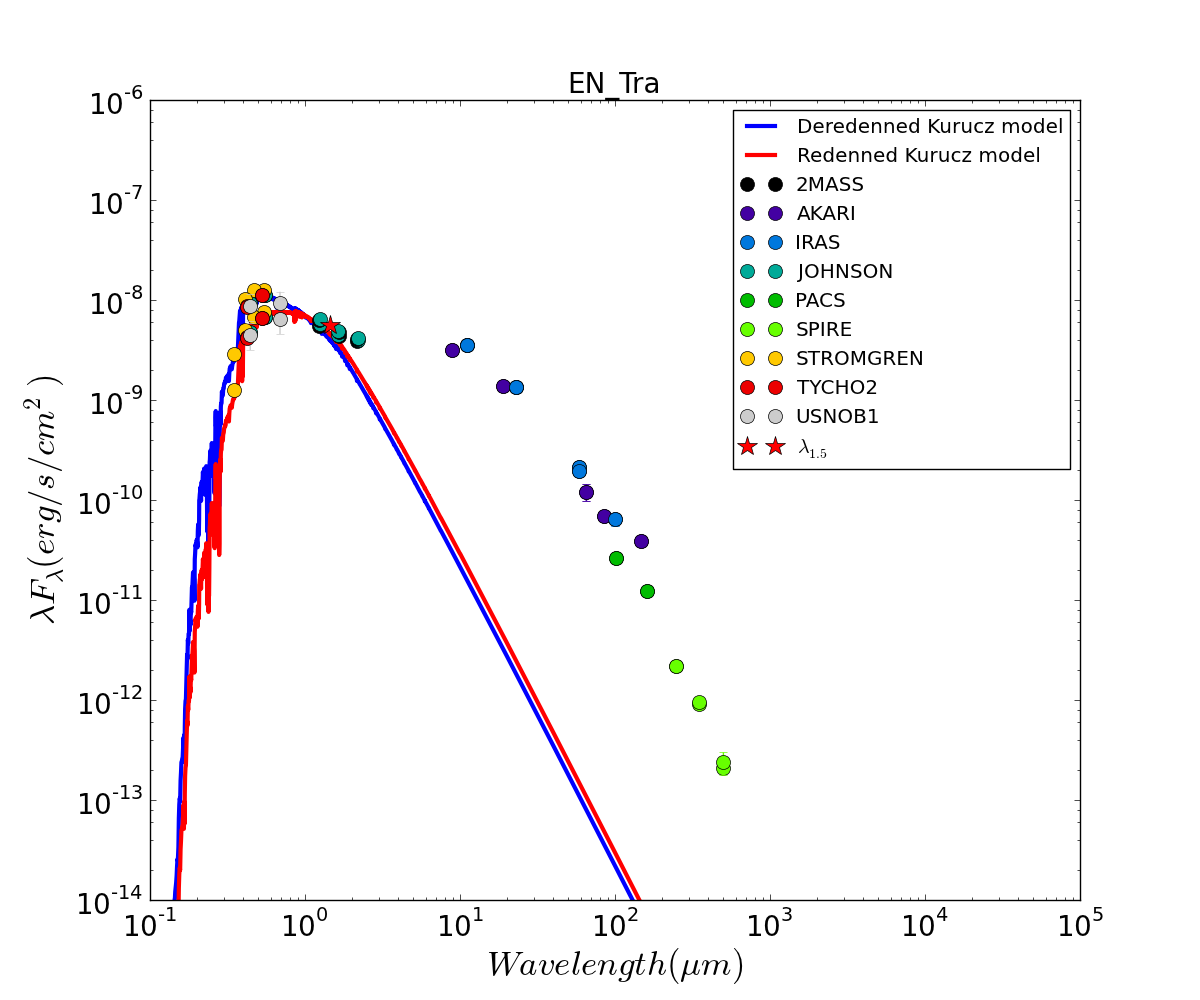}
    \caption[The SED of 89 Her as an example of post-AGB binary]
      {The SED of 89\~Her \citep{hillen13} as an example of an optically bright post-AGB binary. }
    \label{VanWinckel1Fig:sed}
     \end{figure}

While the first binary post-AGB stars were serendipitously detected \citep[e.g.][]{waelkens96} it turned out that their SEDs had some common but distinct properties. This is illustrated in Figure~\ref{VanWinckel1Fig:sed} where a typical SED of a post-AGB binary is displayed. The energetics display a clear near-infrared
excess by thermal emission of dust,  indicating that this circumstellar dust must be close to the central
star, near sublimation temperature. The peak of the dust excess is around 10 $\mu$m and at the long wavelength tail, the spectral index follows the Rayleigh jeans slope up to submm wavelengths. It is now well established that 
these features in the SED indicate the presence of a stable compact disc in the system (see Section~\ref{sect:discs}). We call these type of SEDs, {\sl disc-type SEDs}.

These specific SED characteristics, in combination with the availability of all-sky infrared surveys of IRAS, WISE and AKARI allowed for efficient searches for similar Galactic systems \citep[e.g.][]{deruyter06, gezer15} and in total about 85 Galactic such sources have been identified by now. These SEDs are in clear contrast to the ones of optically bright post-AGB stars with detached shells: the shell of gas and dust, which was expelled when the object was still on the AGB, is cooling and expanding resulting in an SED without a near-IR excess and a peak in the SED sypically around 30-60 $\mu$m \citep{vanwinckel03}. 

Moreover, the SPITZER surveys in the Large and Small Magellanic Clouds \citep{meixner06,bolatto07,gordon11} 
allowed for a systematic approach in the search for post-AGB stars also in these galaxies. After a photometric selection \citep{vanaarle11,kamath14a}, a spectral survey was performed to locate similar evolved sources and to differentiate them from interlopers \citep{,kamath14a, kamath15}. The objects with distinct near-IR excess represent more than half of the population of optically bright post-AGB
stars. Objects with similar SEDs also appear at lower luminosities, indicating that the central evolved star is a 
post-Red Giant Branch (post-RGB) stars, rather than post-AGB stars \citep{kamath16}.

We conclude that within the known samples of the rare post-AGB stars, disc SEDs are very frequently detected and this both in the Galaxy and the Magellanic clouds.

\section{Orbital Properties} \label{sect:orbits}

\begin{figure}
    \includegraphics[width=0.7\textwidth]{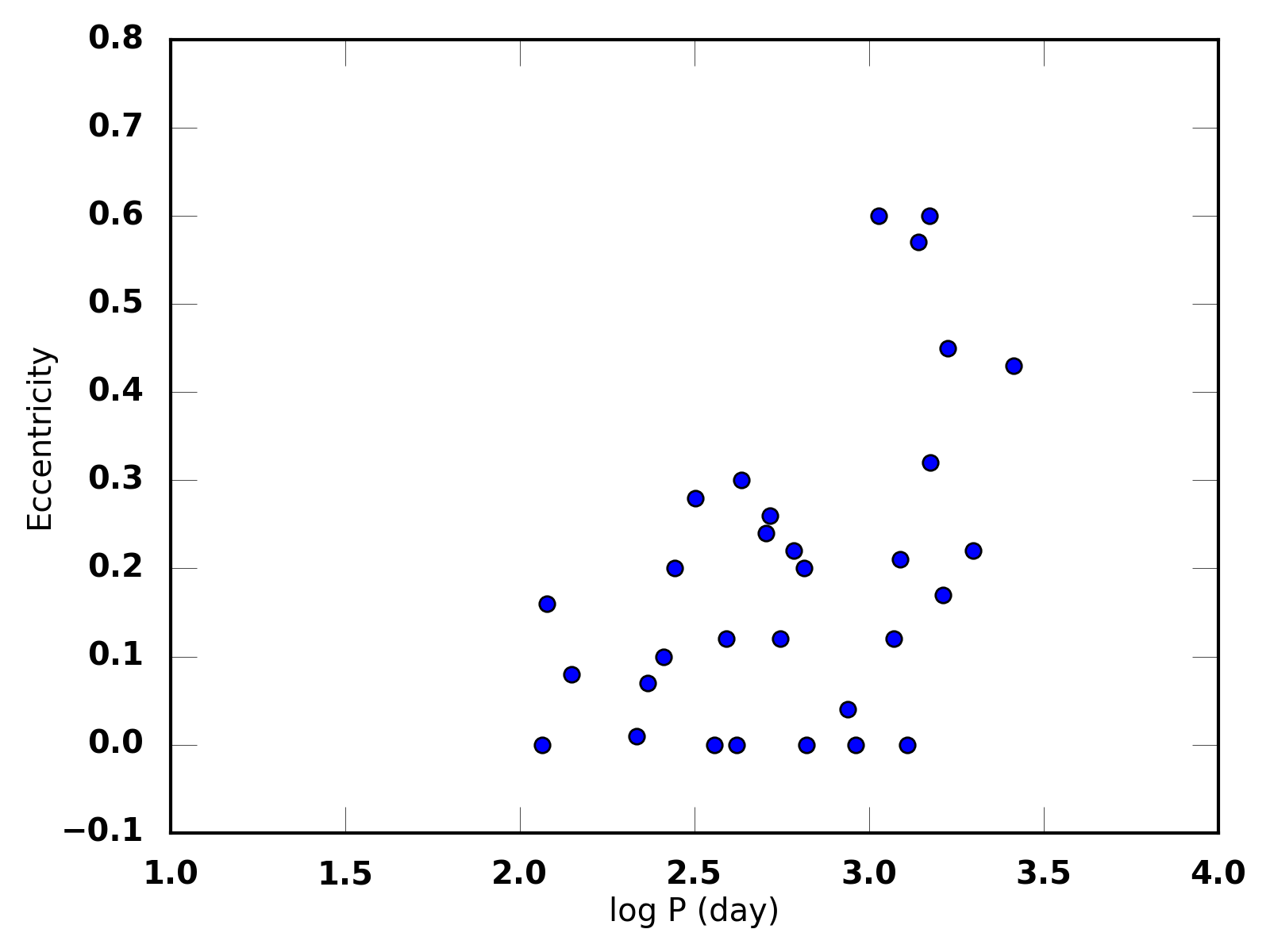}
    \caption[The e-log(P) diagram of post-AGB binaries]
      {The e-log(P) diagram of post-AGB binaries. }
    \label{VanWinckel2Fig:elogp}
     \end{figure}

The binary nature of individual objects \citep{vanwinckel95, waelkens96, gonzalez96,maas02} inspired a systematic monitoring programme of Galactic evolved sources using our own
HERMES spectrograph \citep{raskin11} mounted on the Flemish 1.2m Mercator telescope.
This has resulted in the discovery of many evolved binaries with unexpected periods between 100 and 2000 days 
within the sample of sources with disc SEDs \citep[e.g.][]{vanwinckel09,gorlova12,gezer15,manick17}. The $a \sin i$ is typically $\sim$ 1 AU. The luminous evolved component has a likely unevolved companion with a very minor contribution to the
energy budget \citep[e.g.][]{vanwinckel09}. The Figure~\ref{VanWinckel2Fig:elogp} is illustrative of this. One of the most striking results is a clear lack of spiralled-in systems at short periods in the order of days. This is not a observational bias: these would be the most easy to detect in radial velocity monitoring programmes. Another striking observation is the rare occurrence of zero-eccentricity systems as many objects have significant eccentricities.

The global picture that emerges from the orbital elements is that the star evolved in a system which is too small to accommodate a full grown RGB or AGB star. During a poorly
understood phase of strong interaction, the system did not suffer a dramatic spiral in. The observed orbits (see Figure~\ref{VanWinckel2Fig:elogp}) show that the common envelope was either very rapidly
expelled or somehow avoided. We postulate that this binary interaction phase truncated the AGB or RGB evolution. All the $\sim$ 30 orbits found till now and their non-zero eccentricities fall
in a period regime {\sl not } predicted by standard binary channels. Moreover, the tidal circularisation was not effective and most objects have distinct eccentric orbits. Eccentricity pumping mechanisms need to be invoked to explain the eccentricities. As the dust sublimation
radii for these sources are well beyond the orbit, the dusty discs are circumbinary.

Finding binaries among post-AGB stars with an expanding shell (shell-sources), on the contrary, turned out to be a cumboursome challenge: despite radial-velocity monitoring efforts for decades, confirmed spectroscopic binaries were not detected yet in post-AGB stars with such detached shells which puts strong constraints on the companion mass and/or possible orbits \citep{hrivnak11, hrivnak17}. Interestingly, these detached shells are often resolved and show bipolar and even multipolar geometries \citep{hrivnak17}, which are often associated with binary interaction physics. If at all, these shell objects are likely very wide systems.

Spectroscopic binaries in post-AGB stars are uniquely connected to their SED properties: almost all binaries known till now have disc-SEDs. Very few exceptions exist and these have a very small dust excess which seems to be the relic of the circumbinary disc.
          
\section{Discs: Resolved from the inner Edge to the outer Radius} \label{sect:discs}

\subsection{Near and Mid-IR Interferometry}

It is now well established that this type of disc-SED, is a good observational proxy for the presence of a stable compact
circumbinary disc. As the infrared emission is compact and the objects are typically at a distance of several kpc, interferometric techniques are needed to resolve them. Applications of high-spatial-resolution techniques with limited uv-coverage \citep{deroo06,deroo07,hillen13,hillen14,hillen15,hillen17} are unveiling the very compact infrared emission regions. There is by now ample evidence that these circumbinary discs show many similarities with the passive protoplanetary discs (PPDs) around young stars \citep[e.g.][]{dullemond10,menu15}.

The spatial information obtained by interferometric experiments are combined with detailed physical modelling via radiative transfer calculations, in which the transport of radiation in three
dimensions is treated self-consistently.  The vertical scale of the disc is
computed under the assumption of hydrostatic equilibrium. For the opacity calculations, the opacities of silicates are typically used \citep{min07} with a variable size distribution of the grains. Turbulent mixing and vertical settling of grains can be included self-consistently \citep{mulders12}, as it strongly affects the structure of the disc, and therefore the observables. The good agreement between the best models and the interferometric observations in these references above gives strong support to the physical interpretation of the circumstellar material: the structure is that of a dusty settled disc. The best-fit grain size distribution shows that grain growth is significant. Sub-micron-sized particles hardly
contribute to the total opacity in this disc, while mm-sized grains need to be included to fit the sub-mm fluxes. 

\begin{figure}
    \includegraphics[width=0.5 \textwidth]{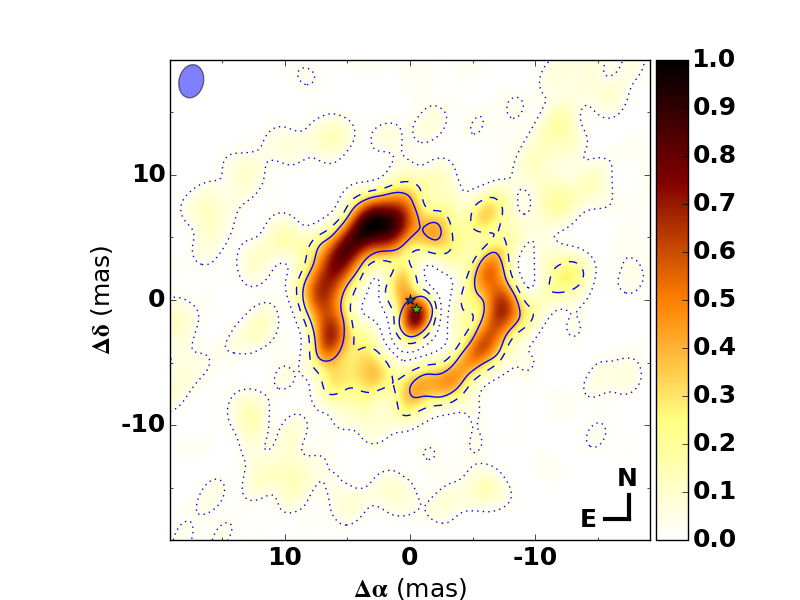}
    \includegraphics[width=0.4 \textwidth]{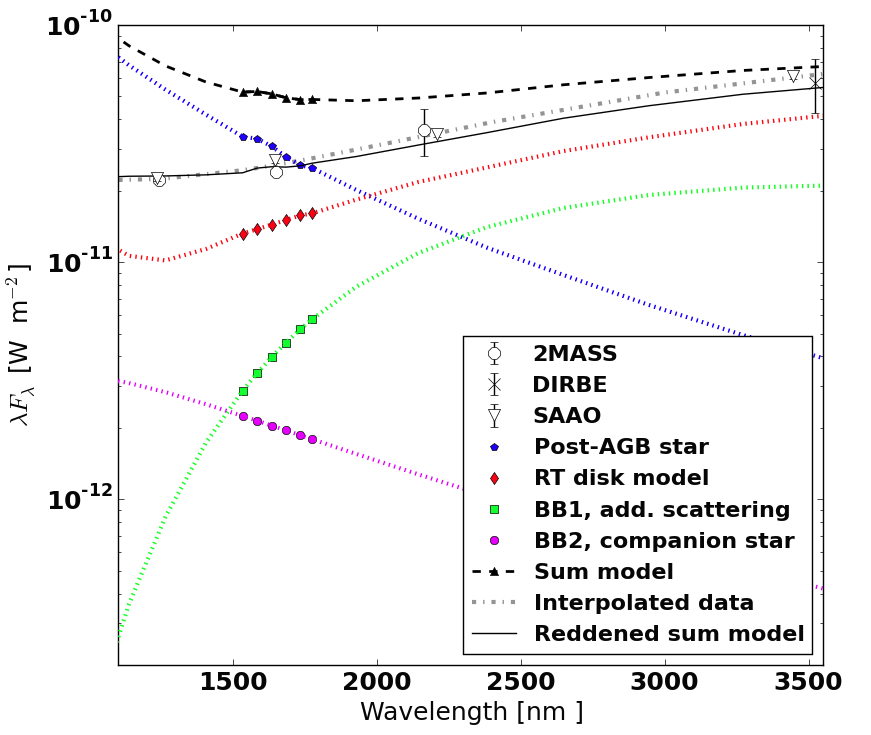}
    \caption[The interferometric image in the H-band of IRAS08544-4431.]
      {The interferometric image in the H-band of IRAS08544-4431 on the left. The central luminous post-AGB star is removed for clarity (blue star). The inner rim of the circumbinary disc is well resolved and the off-centre flux contribution is interpreted as coming from the accretion disc around the companion \citep{hillen16a}. On the right, the dissection of the SED is shown with all spatially resolved contributing components.}
    \label{VanWinckel3Fig:imaging}
     \end{figure}

A first interferometric imaging experiment, which is possible provided the UV-plane is covered extensively, was presented in \cite{hillen16a}.
Thanks to the unprecedented spatial resolution, all important building blocks (circumbinary disc, central binary, accretion disc around the companion) could be spatially resolved and their flux contribution could be quantified which is illustrated in Figure~\ref{VanWinckel3Fig:imaging}. The most apparent is the inner rim of the circumbinary disc, which is indeed at the sublimation radius of the star. For clarity the central bright star, which is unresolved, is removed from the image.  The companion itself is too faint to contribute significantly to the total H-band flux, but it is the accretion disc around the companion which is resolved as an additional contributor to the center of the image, slightly displaced from the luminous post-AGB star. The physical model of the circumbinary disc assumes a settled disc in hydrostatic equilibrium, with an axisymmetric surface density. While the model fits many (interferometric) observables, the real data shows more structure.

Just like for pre-main-sequence stars across a wide range of mass and luminosity \citep{dullemond10,kraus10}, spectro-interferometry proves very powerful to study the physics of settled discs.

\subsection{CO Interferometry}

The most conclusive evidence for stability comes from the detection of resolved Keplerian velocity fields in three systems so far: the Red Rectangle\index{Red Rectangle}, AC Her\index{AC Her} and IW Car\index{TW Car}. These were spatially resolved in CO using the ALMA and the Plateau De Bure interferometers \citep{bujarrabal13a,bujarrabal15,bujarrabal17a}. Moreover, for 89\,Her \index{89 Her} the inner CO disc is found to be not resolved but likely in Keplerian rotation as well \citep{bujarrabal07}.

In addition to the Keplerian velocity, a slowly expanding component was resolved in the maps of the Red Rectangle, 89\,Her and IW Car, showing a bipolar low-velocity
outflow. Given its structure and kinematics, these outflows likely originate from the disc.
The total angular momentum of the disc of the Red Rectangle is significant and suggests that, if the orbit is the source of angular momentum, the binary should have shrunk considerably \citep{bujarrabal16}.
By comparing the total mass of the CO disc with the mass-loss as detected in CO, the estimated lifetimes are between $\sim$ 8000 and 10000 yr \citep{bujarrabal16,bujarrabal17a}.

Single-dish observations
confirm that rotation must be widespread among other disc sources as well: the CO rotational lines are very narrow indicating that Keplerian rotation rather than expansion is detected in all the sources of the survey \citep{bujarrabal13a}. Clearly more systems need to be investigated in detail.

\subsection{Longevity}

The discs have other distinct observational features which are interpreted as indicators of longevity.
Infrared spectroscopic observations of dust features show that strong dust-grain processing is needed to explain the profiles. The processing leads to a high degree of crystallinity in the silicate bands
\citep[e.g.][]{molster02c,gielen07,gielen11}. 

It is remarkable that in all systems, Galactic and from the Maggelanic clouds alike, the dust is oxygen rich and Mg-rich endmembers of the silicates prevail. Aditionally, some objects show a carbon-rich component like the Extended Red Emission (ERE) in the Red Rectangle \citep[e.g.][]{cohen04} and PAH emission in a minority of objects. 

Another indicator for longevity, is the presence of large grains. In order too explain the contrast and width of the silicate bands \citep[e.g.][]{gielen07,gielen11,arneson17}, micron sized grains are needed. The presence of even larger grains, to the cm regime, are needed to explain the spectral index at sub-mm wavelengths \citep[e.g.][]{jura97,deruyter05,gielen07, hillen14, hillen15}.

\section{Outflows and Jets} \label{sect:jets}

\begin{figure}
    \includegraphics[width=0.9\textwidth]{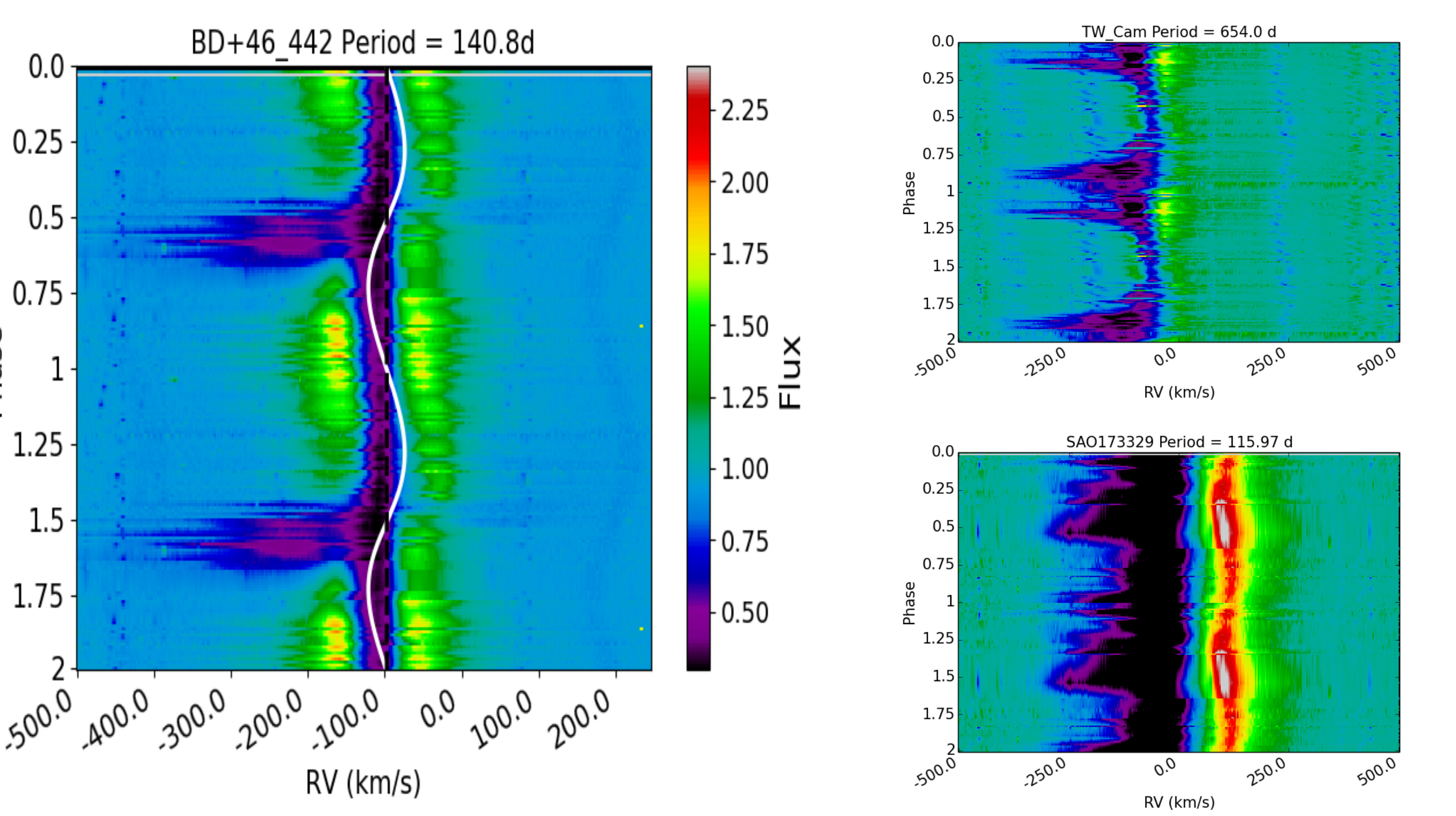}
    \caption[jets]
      {Dynamic H-alpha profiles folded on the orbital period. On the left, the data of BD+46$^{\circ}$442 \citep{bollen17} folded on its 140.7 days orbital period. The full line represents the movement of the primary. The absorbtion at high velocity appears when the companion is in front of the primary (phase 0.5). On the right two other illustrative examples. At the top, the H$_{\alpha}$ profiles of the RV\,Tauri pulsator TW Cam are displayed, folded on the orbital period of 654 days. The pulsations are seen in the photospheric lines  \citep{manick17}. The absorption component appears when the companion is in front (phase 1.0). The bottom shows the profiles of SAO173329 with an orbital period of 116 days. Here we look almost pole-on which makes that the high velocity outflow is always in the line-of-sight to the primary. }
    \label{VanWinckel4Fig:halpha}
     \end{figure}

Another important structural component is detected only indirectly. This is best illustrated by showing
orbital phase-resolved high-resolution spectra (Figure~\ref{VanWinckel4Fig:halpha}). the H$_\alpha$ profile turns into a P-Cygni profile only when the unseen companion is in front of the luminous primary which is the post-AGB star. The interpretation of this behaviour is that a high velocity outflow or jet, originates around the companion and is only detected when continuum photons of the primary are scattered out of the line-of-sight \citep{thomas13, gorlova12, gorlova15}. When a good phase coverage is reached, the wide opening angle as well as the latitudinally dependent velocity law can be deduced from the spectra \citep{thomas13,bollen17}. For the objects which are well studied, the outflow velocity is larger in the polar direction, with a high-velocity and low density on the jet axes, and a lower velocity, denser outflow at the conical edge.
The deprojected velocity corresponds to the escape velocity of a main sequence star, rather than from a compact object.

The jets are not strongly collimated, but are launched likely from a circum-companion accretion disc. This disc is seen outside conjunction via the shell-profiles of the H$_{\alpha}$ emission, which can be used to map in velocity space, via Doppler tomography \citep{bollen17}. Whether the circum-companion accretion disc is fed by direct mass transfer from the primary, or via accretion from the circumbinary disc is not constrained yet.

While modelled and published only in a limited number of systems till now, similar orbital phase-dependent Halpha profiles are now commonly observed in post-AGB binaries (see Figure~\ref{VanWinckel4Fig:halpha}). When a system is seen nearly pole-on, the P-Cygni line-profile is seen at all orbital phases as the wide jet is always in the line-of-sight. Post-AGB binaries provide ideal test beds to study jet formation and launching mechanisms over a wide range of orbital periods and mass accretion regimes.

\section{Feedback from Circumbinary Discs} \label{sect:depletion}

\begin{figure}
    \includegraphics[width=0.7\textwidth]{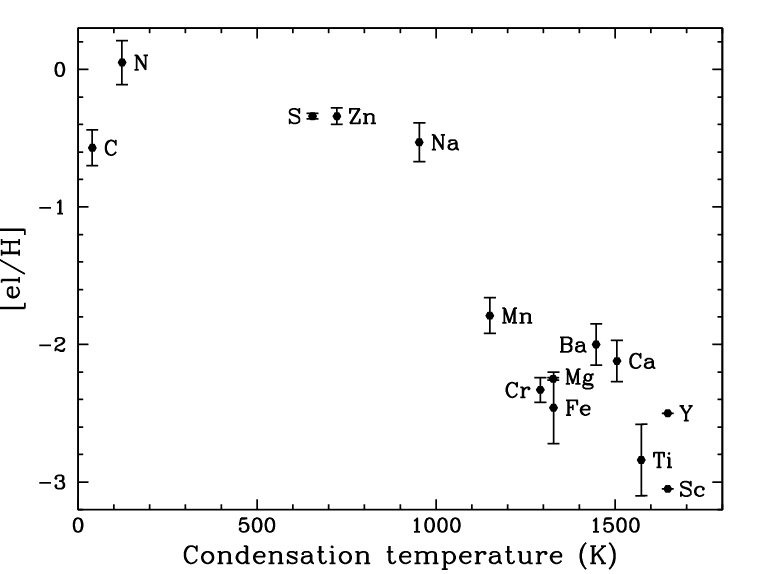}
    \caption[Depletion]
      {A depleted photosphere has a chemical anomaly: the photopheric elemental abundances correlate with the dust condenstationtemperatures. (Figure give the abundances of MACHO82.8405.15 adapted from \citep{gielen09b}). }
    \label{VanWinckel5Fig:depletion}
     \end{figure}

Many post-AGB stars display a chemical anomaly in their photosphere called 'depletion' \citep{vanwinckel03,rao14,venn14}.
The abundance trends in such atmospheres resemble the gas phase of the interstellar medium: refractory elements 
are underabundant, while volatiles retain their original abundances (see Figure~\ref{VanWinckel5Fig:depletion}). 
Good tracers of depletion are the [S/Ti] and [Zn/Ti] or [Zn/Fe] abundance ratios. The $s$-process elements are also refractory \citep{lodders03} so the eventual enrichment of AGB nucleosynthesis products is masked by the depletion process.

The process to acquire this chemical anomaly is not completely understood
yet. Dust formation leads to a chemical fractionation in the circumstellar environment. To obtain a depleted photosphere,
the radiation pressure on circumstellar dust grains must fractionate the dust from the gas. The cleaned gas is then re-accreted
on to the stellar surface.

Such depletion patterns are intimately linked
to the presence of a disc around the post-AGB star, and indicate that gas, cleaned from the refractory elements by dust formation, is accreted to the primary \citep[e.g.][and references therein]{giridhar05,gezer15}. The depletion process can be very efficient and in some case the original Fe content is reduced to [Fe/H]=$-$3 to $-$4., while the non-refractory elements like S and Zn retained their original abundances \citep{vanwinckel95,venn14}. Depleted atmospheres are also found
in post-AGB stars in the LMC and SMC \citep{reyniers07b, gielen09b}.

The longevity of the disc \citep{waters92},  seem to be a needed but not sufficient condition for the process to occur. As shown in a more systematic comparison in \cite{gezer15}, depleted photospheres are striclty limited to objects with a disc. For some exceptions, there is good observational evidence of a former disc. 
If the accretion of mass is similar or larger than the envelope reduction due to external mass-loss and core-growth, the post-AGB lifetime will be lengthened.

\section{Conclusions} \label{sect:conclusion}

The specific SED charateristics of established binary post-AGB stars, allow for a very efficient search for similar disc-SEDs among the evolved optically bright stars. 
Moreover, detailed and long-term radial velocity monitoring programmes have shown that the central stars with disc-SEDs are indeed binaries with  typical orbital periods between 100 and 2000 days (Figure~\ref{VanWinckel2Fig:elogp}), which are orbits least predicted by stellar binary evolution models.  No spiralled-in systems are found, despite the orbital sizes which are too small to accomodate an AGB star. While spiralled-in systems are commonly observed in central stars of PNe \citep[e.g.][]{jones17}, post-AGB stars with discs evolve on tracks which avoid spiralling-in. Post-AGB binaries are by no means the only evolved binaries with orbits that are not well understood (see e.g. chapter by Pols).

The discs around evolved binaries are scaled-up versions of protoplanetary discs, because the ratio of luminosity versus mass is much higher. They have a physically larger dust-free inner region of several astronomical units. It is remarkable that a hydrostatic model, dominated by dust sublimation physics can represent the morphology and energetics of the circumbinary disc even better than it does for discs around young stars, for which these models were originally developed \citep{hillen17}. Distances are typically poorly constrained, except for the disc sources in the LMC and SMC but for the latter, orbital properties are as yet not derived \citep{kamath14a,kamath15}. This does hamper the comparison with binary models. Constraining the distances will become possible when GAIA accumulates enough data that enables differentiation between parallax movements and orbital movements.

There is increasing evidence that the presence of a circumbinary disc has a lasting effect on the dynamical
evolution of the binary orbit, as resonant coupling between the binary and the disc can pump-up the eccentricity
\citep{artymowicz91,lubow10,dermine13,vos15}. This provides a way to explain
the commonly observed large eccentricities. The integrated effect depends strongly on the mass and longevity of the disc, as well as on the radial mass distribution within the disc. Quantification of this eccentricity-pumping effect is hence
only possible when the evolution and timescale of the discs are understood. Moreover the total angular
momentum stored in the discs are, so far, determined in a few systems only \citep{bujarrabal17a}.

Jets are seen as a major shaping agent in PNe \citep{balick02, demarco17} and are thought to shape the multipolar and often complex reflection nebulae in proto-planetary stars \citep{sahai07}. Grazing jets \citep{soker17,shiber17} are potentially important for avoiding common-envelope phase in many binary systems. Despite their importance, jet physics and launching mechanisms are still poorly understood. The optically bright post-AGB binaries offer ideal test beds for jets and their creation processes. The circumcompanion accretion discs where they are launched can be fed either by the primary, or by the circumbinary dusty disc or both. The reflection nebulae, which are shell-sources, are often very multi-polar and structured, but for these, the monitoring data complies only with either single stars, very wide systems, or very low-mass companions \citep{hrivnak17}. These shell systems are very different than the objects with disc-SEDs.

We conclude that stable circumbinary discs are integral component of many post-AGB binaries and are likely playing a lead role in the evolution of the systems. We can postulate the disc creation, evolution and evaporation play a lead role in all systems with a former AGB star. Detailed studies of the structure and evolution of these discs and their impact on the orbital evolution are therefore badly needed.

\section{Acknowledgements}
 It is a pleasure to acknowledge the many colleagues, post-docs and PhD students which contributed recently to the research on post-AGB binaries: Christoffel Waelkens, Devika Kamath, Peter Wood, Valentin Bujarrabal, Orsola De Marco, Michel Hillen, Alain Jorissen, Sophie van Eck, Lionel Siess, Onno Pols, Gijs Nelemans, Bruce Hrivnak, Brent Miszalski, Rajeev Manick, Ana Escorza, Shreeya Shetye, Glenn-Micheal Oomen, Dylan Bollen, Ilknur Gezer and Joonas Sario.

  \bibliographystyle{cambridgeauthordate}


    
\end{document}